\documentclass[final]{svjour_mod} 
\usepackage{graphicx}
\usepackage{rotating}
\usepackage{amssymb}
\usepackage{mathptmx}
\usepackage{amsfonts}
\usepackage{amsmath}
\usepackage{bm}
\usepackage[numbers]{natbib} 

\makeatletter
\journalname{Journal of Low Temperature Physics}

\bibpunct{}{}{,}{s}{}{,}

\begin{document}
\newcommand{\hdblarrow}{H\makebox[0.9ex][l]{$\downdownarrows$}-}

\newcommand{\vs}{\textbf{v}_{\mathrm s}}
\newcommand{\vn}{\textbf{v}_{\mathrm n}}
\newcommand{\vl}{\textbf{v}_{\mathrm L}}
\newcommand{\he}[1]{$^{#1}$He}
\def\Oms{\Omega_{\rm s}}

\newcommand{\ud}{\mathrm}
\newcommand{\tc}{\,T_{\ud c}}


\newcommand{\eq}[1]{(\ref{#1})}
\newcommand{\Eq}[1]{Eq.~(\ref{#1})}
\newcommand{\Eqs}[1]{Eqs.~(\ref{#1})}
\newcommand{\Fig}[1]{Fig.~\ref{#1}}
\newcommand{\Figs}[1]{Figs.~\ref{#1}}
\newcommand{\Sec}[1]{Sec.~\ref{#1}}
\newcommand{\Secs}[1]{Secs.~\ref{#1}}
\newcommand{\Ref}[1]{Ref.~\cite{#1}}
\newcommand{\Refs}[1]{Refs.~\cite{#1}}

\def\be{\begin{equation}}\def\ee{\end{equation}}
\def\bea{\begin{eqnarray}}\def\eea{\end{eqnarray}}
\def\bse{\begin{subequations}}\def\ese{\end{subequations}}
\newcommand{\BE}[1]{\begin{equation}\label{#1}}
\newcommand{\BEA}[1]{\begin{eqnarray}\label{#1}}
\newcommand{\BSE}[1]{\begin{subequations}\label{#1}}

\let \nn  \nonumber  \newcommand{\br}{\\ \nn}
\newcommand{\BR}[1]{\\ \label{#1}}
\def\hf{\frac{1}{2}}
\let \= \equiv \let\*\cdot \let\~\widetilde \let\^\widehat \let\-\overline
\let\p\partial \def\pp {\perp} \def\pl {\parallel}
\def\ort#1{\^{\bf{#1}}}
\def\Trans{^{\scriptscriptstyle{\rm T}}}
\def\x{\ort x} \def\y{\ort y} \def\z{\ort z}
 \def\bn{\bm\nabla} \def\1{\bm1} \def\Tr{{\rm Tr}}
\def\Re{\mbox{  Re}}
\def\<{\left\langle}    \def\>{\right\rangle}
\def\({\left(}          \def\){\right)}
 \def \[ {\left [} \def \] {\right ]}

\renewcommand{\a}{\alpha}\renewcommand{\b}{\beta}\newcommand{\g}{\gamma}
\newcommand{\G} {\Gamma}\renewcommand{\d}{\delta}
\newcommand{\D}{\Delta}\newcommand{\e}{\epsilon}\newcommand{\ve}{\varepsilon}
\newcommand{\E}{\Epsilon}\renewcommand{\o}{\omega} \renewcommand{\O}{\Omega}
\renewcommand{\L}{\Lambda}\renewcommand{\l}{\lambda}
\renewcommand{\t}{\tau}
\def\r{\rho}\def\k{\kappa}
\def\t{\theta } \def\T{\Theta } \def\s{\sigma} \def\S{\Sigma}

\newcommand{\B}[1]{{\bm{#1}}}
\newcommand{\C}[1]{{\mathcal{#1}}}    
\newcommand{\BC}[1]{\bm{\mathcal{#1}}}
\newcommand{\F}[1]{{\mathfrak{#1}}}
\newcommand{\BF}[1]{{\bm{\F {#1}}}}

\renewcommand{\sb}[1]{_{\text {#1}}}  
\renewcommand{\sp}[1]{^{\text {#1}}}  
\newcommand{\Sp}[1]{^{^{\text {#1}}}} 
\def\Sb#1{_{\scriptscriptstyle\rm{#1}}}
\hyphenation{Vo-lo-vik Kelvin-Helm-holtz}

\title{Thermal Detection of Turbulent and Laminar Dissipation in Vortex Front Motion}

\author{J.J. Hosio$^1$ \and V.B. Eltsov$^1$ \and M. Krusius$^1$}

\institute{1:O.V. Lounasmaa Laboratory,\\ P.O. Box 15100, FI-00076 AALTO, Finland\\
Tel.:+358 50 344 2483  \\ 
\email{jaakko.hosio@aalto.fi}
}

\date{\today}

\maketitle


\begin{abstract}

We report on direct measurements of the energy dissipated in the spin-up of the superfluid component of $^3$He-B. A vortex-free sample is prepared in a cylindrical container, where the normal component rotates at constant angular velocity. At a temperature of 0.20$\tc$, seed vortices are injected into the system using the shear-flow instability at the interface between $^3$He-B and $^3$He-A. These vortices interact and create a turbulent burst, which sets a propagating vortex front into motion. In the following process, the free energy stored in the initial vortex-free state is dissipated leading to the emission of thermal excitations, which we observe with a bolometric measurement. We find that the  turbulent front contains less than the equilibrium number of vortices and that the superfluid behind the front is partially decoupled from the reference frame of the container.
The final equilibrium state is approached in the form of a slow laminar spin-up as demonstrated by the slowly decaying tail of the thermal signal.

\end{abstract}

\section{Introduction}

Turbulence in superfluids is characterized by interactions between quantized vortex lines, their tangling, and reconnections between the vortices. In a container rotating at constant angular velocity, the turbulent motion of a superfluid can be triggered by applying a sudden sufficiently strong flow perturbation. Eventually the turbulence will decay as the superfluid achieves equilibrium solid-body rotation. In some cases this is achieved via a process which displays steady-state turbulent motion. An example is a turbulent vortex front which propagates at a steady velocity in $^3$He-B along a long rotating tube.\cite{front1} In $^3$He-B, the normal component is practically always clamped to corotation with the container. At low temperatures, this happens such that the dilute gas of of ballistic thermal quasiparticles follows the rotation of the container via diffusive scattering from the container walls.

At finite temperatures, the quantized vortex lines mediate the interaction between the superfluid and the normal components via the mutual-friction force. This arises from the scattering of the thermal excitations from the vortices moving with respect to the rest frame of the normal fluid. These interactions provide dissipation, couple the superfluid component to corotation with the normal fluid, and thus, suppress turbulence provided the normal fluid density is high enough. The equation of motion for the vortex-line segment is obtained from the balance of the Magnus and the mutual-friction forces. The velocity of a vortex-line element in terms of the superfluid counterflow velocity $\vn -  \vs$, where $\vs$ and $\vn$ are the local velocities of the superfluid and the normal components, is given by\cite{donnelly}
\begin{equation}
\vl=\vs +\a\hat{\textbf s} \times (\vn -  \vs)
-\a' \hat{\textbf s} \times[\hat{\textbf s} \times(\vn - \vs)].
\label{vlin}
\end{equation}
Here $\hat{\textbf s}$ is a unit vector along the vortex line, while $\a$ and $\a'$ are the dissipative and the reactive mutual-friction parameters, which depend both on temperature and pressure.\cite{kopnin,bevan}
The stability of the flow is characterized by the superfluid Reynolds
number $\rm{Re}_{\a}=(1-\a')/\a$. Typically, turbulence in bulk volume
becomes possible if $\rm{Re}_{\a}$ is larger than unity,\cite{volovik04}
while the exact critical value depends on the process and can be much higher.\cite{PRL10}

The dissipation of vortex motion rapidly decreases with decreasing
temperature, as the amount of the normal fluid is reduced exponentially in the
$T\rightarrow 0$ limit. However, reconnection-driven turbulent processes\cite{front1,Manchester} and surface interactions\cite{AB_turb,prec} provide finite dissipation even in the limit of vanishing normal-fluid density. At length scales smaller than the inter-vortex distance, the kinetic energy is believed to be transferred to smaller scales in a cascade of helical deformations of individual vortices called Kelvin waves.\cite{vinen} The energy is dissipated throughout the cascade owing to the mutual friction, and ultimately at the lowest temperatures, the cascade may terminate when Kelvin waves with a very short wavelength induce emission of bulk excitations: phonons in $^4$He,\cite{vinenprb} and fermionic quasiparticles in $^3$He-B.\cite{silaev}

In Ref.\citenum{front2}, we presented the first direct observation of these
quasiparticles:
Turbulence created in a steady-state propagation of a superfluid vortex
front was measured calorimetrically (calorimetric measurements of decaying
turbulence are reported in Ref.\citenum{lanc_nat}). The thermal signal
revealed that in the low-temperature limit, the superfluid component behind the front develops its own rotating reference frame, whose angular velocity is
smaller than that of the container and is decreasing with decreasing
coupling. Here we present an extension on those measurements. We describe
the technical implementations of the bolometric measurements, discuss the
energy balance of the superfluid spin-up and the effect of the triggering
mechanism on the thermal signal, and present a phenomenological model
to analyze the observed thermal records.

\section{Superfluid vortex front}

In a smooth-walled cylindrical container rotating at constant angular velocity $\Omega$, superfluid $^3$He-B can exist forever in the metastable vortex-free state, the so-called Landau state. In our experiments, we first prepare the vortex-free state at the velocity $\Omega$, then inject some vortices to the system, and monitor how they move and interact. The ensuing dynamics of the expanding vortices demonstrate a variety of features in steady-state conditions, depending most prominently on temperature, but also on the rotation velocity and the configuration of the seed vortices.\cite{PLTP_rota}

At high temperatures, above 0.6$\tc$ at 29~bar, the seed vortices expand along the container wall in a laminar fashion with the axial vortex-end velocity \begin{equation}
\label{vlam}
\textbf{v}_{\ud L,z}\approx\a(T)\Omega R,
\end{equation}
obtained from Eq. (\ref{vlin}), and the number of vortices is conserved.\cite{finne} At lower temperatures the injected vortices interact in a turbulent manner and create a large number of new vortices in a burst-like process.\cite{Nature03} The details of this turbulent injection process depend on the seed-vortex configuration and thus on the strength of the flow perturbation.\cite{precursor}

At low temperatures, the expanding vortices form a front, in which the ends of the vortex lines bend to the cylindrical wall of the container. The front precesses azimuthally at a lower angular velocity than the bundle behind it, and consequently, they both are twisted.\cite{twist_prl} The effects of vortex curvature, both collective \cite{twist_jltp,sonin_front} and the single-vortex effects,\cite{karimaki} somewhat reduce the propagation velocity $V_{\rm f}$ of the front from that in Eq. (\ref{vlam}) at temperatures down to $T\approx0.45\tc$. At still lower temperatures turbulent processes provide extra dissipation in addition to that from mutual friction and make the propagation faster than in Eq. (\ref{vlam}).\cite{front1} The front is one of the rare examples of steady-state turbulent vortex motion in the bulk volume that can be studied in experiments in the $T\rightarrow 0$ limit.

\section{Energy balance of superfluid spin-up}

In the simplest model of front propagation, all the dissipation is concentrated in the thin core of the front, and the bundle behind the front contains the equilibrium number of vortices $(2\O/\kappa)\pi R^2$, where $\kappa=0.066$\,mm$^2$/s is the circulation quantum. Thus, the core separates the non-rotating superfluid with ${\mathbf v}_{\rm s} = 0$  from that in almost equilibrium rotation with ${\mathbf v}_{\rm s} \approx {\mathbf v}_{\rm n} = \Omega\hat{\mathbf{z}}\times{\mathbf r}$ behind the front. To clarify how much heat is released during the axially propagating spin-up process, we write the energy balance per unit length of the cylinder
\begin{equation}\label{bal}
Q + (E_{\rm fin} - E_{\rm ini}) = A\,,
\end{equation}
where $Q$ is the released heat, $E_{\rm ini}$ and $E_{\rm fin}$ are the initial and the final values of the internal energy of the superfluid, and $A$ is the work performed by external forces.  The energy balance can be analyzed either in the laboratory frame or in the rotating reference frame.

\paragraph{Laboratory frame}
In the laboratory frame the relevant energy $E$ is the kinetic energy of the superfluid per unit length. Thus $E_{\rm ini} = 0$ and $E_{\rm fin} = E\sb{kin}$. For solid-body rotation at $\Omega$ this is given by
\begin{equation}\label{Ekin}
E_{\rm kin} = 2\pi \rho_{\rm s} \int_0^R \frac{ (\Omega r)^2}{2} r dr =
\frac{\pi\rho_{\rm s}}{4}R^4\Omega^2 = E_0 \ .
\end{equation}
Here we neglect the thin vortex-free annulus next to the cylinder wall. The work performed by the motor driving the rotation is
\begin{equation}\label{work}
A = \int M d\varphi = \int M \Omega dt\,,
\end{equation}
where $M$ is the torque and $\varphi = \Omega t$ is the rotation angle of the
cylinder. Since the front propagates at constant $\Omega$, the integral in Eq.~(\ref{work}) is
\begin{equation}
A = \Omega\int M dt = \Omega \, \Delta L\,,
\end{equation}
where $\Delta L$ is the change in angular momentum across the front. The angular momentum of the rotating superfluid is given by
\begin{equation}\label{M}
L = 2\pi \rho_{\rm s}  \int_0^R  (\Omega r) r \cdot  r dr = \frac{\pi\rho_{\rm s}}{2}R^4\Omega \equiv 2\, \frac{E_0}\Omega\ .
\end{equation}
Thus $A=2\, E_0$ and the energy balance~\eq{bal} takes the form $Q + E_0 = 2 E_0 $. This gives the expected result
\begin{equation}\label{res1}
Q =  E_0 = E\sb{kin}=\frac{\pi\rho_{\rm s}}{4}R^4\Omega^2 \,.
\end{equation}

\paragraph{Rotating frame}
In the rotating frame the analysis of \Eq{bal}  is even simpler, because the motor work $A=0$, i.e., the motor which rotates the cylinder, does not produce any work. The reason is that the walls with the clamped normal component do not move, the Coriolis force produces no work, and the centrifugal force is balanced by the pressure gradient. The relevant free energy is $F=E\sb {kin}- \Omega L$, where $E\sb {kin}$,  $\Omega$ and  $L$ are taken in the laboratory frame. Therefore, $F_{\rm ini} = 0$ and $F_{\rm fin} = E_0 - 2E_0= -E_0$. The energy balance is given by $Q - E_0 = 0$ which is the same as  \Eq{res1}.

As expected, the analysis in the laboratory and rotating frames gives the same answer, $Q = E\sb {kin}$. This result is valid only because $\O$ is constant during our measurement and the normal component is corotating with the container.
By measuring the velocity of the turbulent front, the dissipation is obtained as
\begin{equation}\label{dissi1}
dQ/dt = \dot{Q} = \frac{\pi\rho_{\rm s}}{4}R^4\Omega^2 V_{\rm f} \,.
\end{equation}
A measurement of $\dot{Q}$ provides a means to analyze the characteristics
of the front motion. For steady-state front motion \Eq{dissi1} gives the
maximum possible rate of heat-release. In the following we will consider
cases when the number of vortices behind the front is less than the equilibrium number and is modeled by the superfluid in solid-body rotation at angular
velocity $\Oms < \Omega$ such that $\vs = \Oms \hat{\bf z} \times {\bf r}$. In this state $E_{\rm kin} = E_0 (\Oms/\Omega)^2$ and $L = 2 E_0 \Oms /
\Omega^2$. Thus, the free energy in the rotating frame is given by
\begin{equation}
F(\Oms) = E_0 [ (1-\Oms/\Omega)^2 - 1]
\label{freeen}
\end{equation}
and the dissipation rate $\dot{Q} =  -dF/dt$ is reduced compared to Eq.~(\ref{dissi1}).

As can be seen from Eqs.~(\ref{dissi1}) and (\ref{freeen}) the rate of the heat release $\dot Q$ depends strongly on the rotation velocity $\Omega$. On the other hand, the temperature dependence of the signal below about $0.25T_{\rm c}$ is weak since the front velocity appears to saturate at the value of $V_{\rm f} \approx \alpha_\mathrm{eff} \, \Omega R$ where $\alpha_\mathrm{eff} \sim 0.1$  is an approximately temperature-independent parameter generated by the turbulence in the front.\cite{front1}

\section{Experimental techniques}
In the sample setup of Fig.~\ref{trigger}, the motion of vortices and their distribution is monitored with two nuclear-magnetic-resonance (NMR) detector coils. The sample is divided in two sections of \he{3}-B of the same length by a magnetic-field-stabilized layer of \he{3}-A. The two $^3$He-B sample sections above the 0.75\,mm orifice can be organized to rotate around the cylinder axis at constant angular velocity $\O$ in the vortex-free Landau
state.\cite{VorFormAnnih}
\begin{figure}
\begin{center}
\includegraphics[%
  width=0.6\linewidth,
  keepaspectratio]{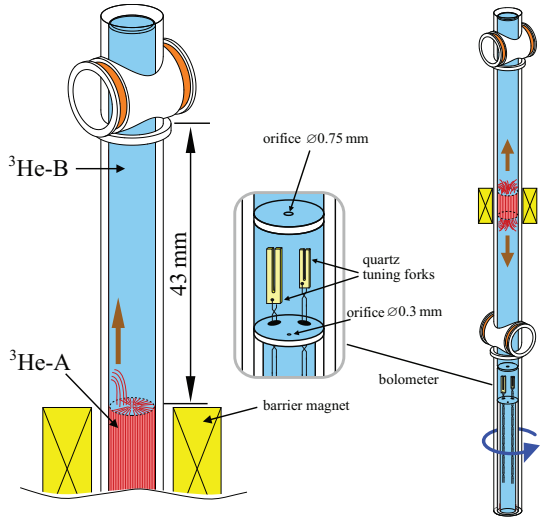}
  \vspace{-5mm}
\end{center}
\caption{(color online) Experimental setup. The front is triggered by manipulating the current in the barrier magnet in the middle of the sample volume. Either the current is decreased to make the interfaces of a pre-existing A-phase layer unstable or increased to create an A-phase barrier layer. Both methods lead to an escape of vortices to the B-phase sections and trigger  turbulent bursts setting both upward and downward front propagation into motion. The front motion in the top section above the 0.75~mm orifice is monitored with two independent continuous-wave NMR detectors. The middle section, which is separated from the heat-exchanger volume with a 0.3~mm pinhole, houses two quartz tuning oscillators which are used for thermometry and for the bolometric measurement of the heat release.}
\label{trigger}
\vspace{-5mm}
\end{figure}

Our thermal measurements of the front propagation are performed at 29\,bar with $\O$ ranging from $0.8$ to $1.2$\,rad/s. The volume above the 0.3\,mm orifice with 5.85~mm inner diameter at $0.20\,T\sb{c}$ acts as a bolometer, an enclosure with a weak thermal link to the outside superfluid $^3$He at much lower temperature $T<0.14\,T\sb{c}$. The thermal response from the vortex motion in the uppermost volume is recorded as a temperature rise inside the bolometer with the quartz tuning fork oscillators in the middle section between the division plates. The lowermost volume, which is in in good thermal contact with the sintered-silver heat exchanger, is filled with an equilibrium number of vortices in the steady state of rotation.

\paragraph{Triggering front motion}
Our method to trigger the front motion relies on the so-called Kelvin-Helmholtz shear flow instability of the A-B interface. \cite{KH_instab}  The instability occurs at a well-defined critical velocity $\Omega\Sb{AB}(T,P)$ which depends on the magnetic stabilization field $H_\mathrm{b} = H\Sb{AB} (T,P)$ and its gradient at the A-B interface. In our thermal measurements $\Omega\Sb{AB}$ is traversed by sweeping $H_\mathrm{b}$ at constant rotation velocity. Prior to the instability event the A-phase slab contains rectilinear vortices which bend onto the A-B interface forming a vortex sheet, in which the vortices run radially to the cylindrical side wall. The instability causes some vortex loops to escape across the interface to the B-phase side. The loops interact and produce a turbulent burst, where a large number of vortices is created in a series of vortex reconnections. In the setup of Fig.~\ref{trigger}, the instability occurs simultaneously at both A-B interfaces setting both upward and downward propagating fronts into motion.

In one set of experiments $\Omega$ is chosen to be initially below  $\Omega\Sb{AB}(H_\mathrm{b})$. The field $H_\mathrm{b}$ is then swept down, whereby $\Omega\Sb{AB}$ decreases, until the instability builds up on the A-B interface. No signal is obtained of the instability event itself, but $\Omega\Sb{AB}$ is reproducible and has been determined accurately from separate measurements. The present temperatures are lower than in the measurements of $\Omega\Sb{AB}$ in Ref.~\citenum{KH_instab}. Nevertheless, our values of $\Omega\Sb{AB}$ lie on straightforward extrapolations of the earlier measurements. Thus the moment, when the front is launched, is identified independently with good precision.

The second means of triggering is to start the experiment with no A-phase layer. When $H_\mathrm{b}$ is swept up until A-phase formation starts, sudden vortex formation is triggered provided that $\Omega > \Omega\Sb{AB}$. The A-phase formation is accompanied by a sharp pulse of cooling which provides a convenient signal of the trigger. The cooling effect is later removed by subtracting the equivalent signal measured in the equilibrium vortex state at the same value of $\Omega$.

\paragraph{Detecting front propagation with NMR}
\begin{figure}
\begin{center}
\includegraphics[%
  width=0.75\linewidth,
  keepaspectratio]{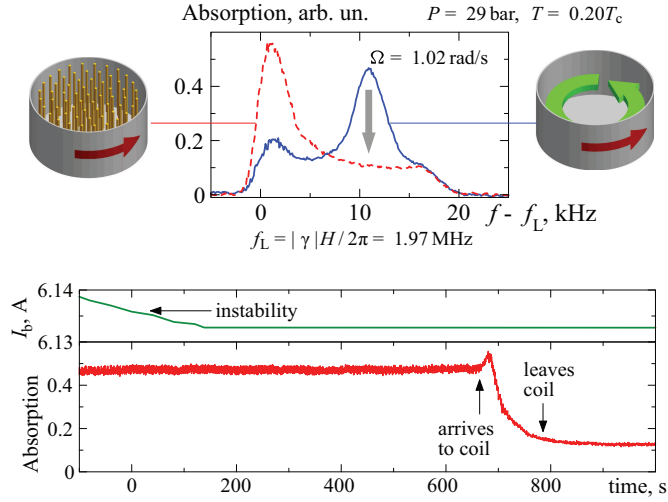}
\end{center}
\caption{(color online) Detection of the front motion with NMR.
(\emph{Top}) NMR absorption signals in constant rotation with and without vortices in the sample. In large vortex-free counterflow the maximum of the absorption is shifted from the Larmor frequency $f_{\ud L}$ to the counterflow peak (solid line). When the sample becomes filled with vortices, the counterflow peak disappears and the absorption is transferred towards the Larmor value (dashed line).
(\emph{Bottom}) Front signal in NMR absorption. A small reduction of the current $I_b$ in the barrier magnet (upper panel) triggers the Kelvin-Helmholtz instability at the A-B interface at $t=0$. The arrival of the front to the detector coil is seen as a drop of the absorption signal, which is monitored at the frequency of the counterflow peak (lower panel).
}
\label{nmrsignal}
\vspace{-5mm}
\end{figure}
The arrival of the front to the end of the sample tube, and thus the velocity of the front, are determined by following the NMR signals of the two detector coils. In the vortex-free counterflow  $\vs=0$ the maximum of the NMR absorption spectrum is shifted from the Larmor frequency $f_{\ud L}$ to the so-called counterflow peak. With $\vn-\vs=0$, i.e., when the sample is filled with an equilibrium number of vortices,  the absorption maximum is close to the  Larmor value. The height of the counterflow peak depends strongly on temperature \cite{deGraaf} but is still sensitive to changes in the amount of counterflow at $0.2\tc$ as demonstrated in the top part of Fig. \ref{nmrsignal}.

In the bottom coil in Fig. \ref{trigger}, the front is detected by tracing
the NMR signal at the counterflow peak. The arrival of the front after
$\sim10$ minutes from the trigger is seen as a decrease of the NMR
absorption as shown in the bottom part of Fig. \ref{nmrsignal}. In the top
coil operating at a lower value of the steady magnetic field
($|\gamma|H/2\pi = 0.87\,$MHz), the front propagation is usually monitored
using a nonlinear NMR mode, based on the condensation of magnons to the
trap, created by the order-parameter texture close to the axis of the
sample. The frequency shift of the magnon-condensate peak from the Larmor
value depends on the trap profile and is a sensitive probe of the local
counterflow.\cite{magnon}

\paragraph{Bolometric measurements of quasiparticle emission}
The energy dissipation is measured as a temperature increase across the
thermal resistance of the bolometer orifice $R_{\rm T} (T)=(d\dot{Q}/dT)^{-1}$. This resistance depends on temperature and on the so-called effective area $A_h$ of the orifice, which can be calibrated using one quartz tuning fork as
heater and another as thermometer. Due to the Andreev reflection from the
flow field created by the vortices in the volume below the bolometer, the
value of $A_h$ depends on $\O$.\cite{andreev_hosio}  Another way to obtain
the power calibration is to measure the thermal time constant of the
bolometer $\tau_{\rm T}=R_{\rm T}C$, where $C$ is the heat capacity of the
$^3$He-B in the bolometer. The uncertainty in the first power-calibration
method arises from the uncertainty in the input-power calibration, while
the temperature calibration of the thermometer has no effect. In the second
method the situation is the opposite. The analysis in this paper is done
using the first (traditional) calibration method. The second method gives
$\sim30\%$ larger values for the dissipated power. We take this difference to represent an upper limit of the absolute uncertainty in the power measurements.

\begin{figure}\sidecaption
\resizebox{0.65\hsize}{!}{\includegraphics*{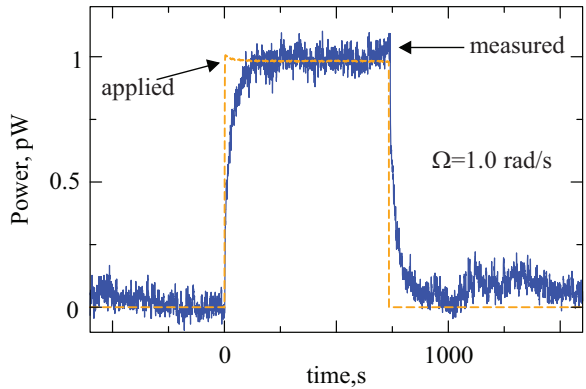}}
\caption{(color online) Calibration of the bolometer. A 1~pW heat pulse shown by the dashed line is fed to the sample with one quartz tuning fork, and the thermal response of the bolometer shown by the solid line is measured with another. The resolution is limited by the noise of the order of 0.1~pW, by somewhat larger fluctuations from the mean value at $\O=1$~rad/s, and by the bolometer response time $\tau_{T}\approx 25\,$s.}
\label{bolcal}
\end{figure}
In Fig. \ref{bolcal}, an example of the bolometer response to a 1~pW heat
pulse is shown. The noise level of the signal is about 0.1~pW , while
fluctuations in its background level are 2--3 times larger at
$\O=1$~rad/s. The thermal relaxation time is $\tau_{T}\approx 25\,$s, while
temperature equilibrium within the bolometer volume is established in a
couple of seconds. These times should be compared to the time which the
fronts propagate from the A-B interfaces to the cylinder end plates, which
is about $500\,$s, or to the total time 2000--3500\,s, over which the
thermal signal is recorded.

The thermometer fork is calibrated at $T=0.33\tc$ against the $^3$He melting curve thermometer, which is mounted on the copper frame of the $^3$He heat-exchanger volume. Lower temperatures are read by assuming the resonance width of the fork oscillator to be
proportional to $\exp(-\Delta/T)$, where the superfluid energy gap
$\Delta\approx1.968\,T_{\rm c}$ at 29~bar pressure is found by linear-in-density interpolation between the BCS value at zero pressure and the value measured in Ref.~\citenum{todo} at the melting pressure.

\section{Thermal measurements}

Figure \ref{signal} shows an example of the thermal response of the bolometer after triggering the front motion at $t=0$. The integrated absorbed energy of the cooling spike from the A-phase creation used as a trigger can be compared to the latent heat of the A-B transition measured at lower pressures by Bartkowiak and coworkers.\cite{PhysRevLett.83.3462} Based on the extrapolation of their measurements to the pressure of 29 bar, the volume of the created A phase should be about 60~mm$^3$. This is in reasonable correspondence with the calculated shape of the two A-B interfaces when they are initially formed.\cite{Abar_shape} Direct measurement of the latent heat from the cooling spike is complicated by the additional dissipation connected with the quickly growing A phase. \cite{PhysRevLett.85.4321}

The measured total heat release $Q$ in Fig. \ref{signal} is less than in Eq.~(\ref{res1}), roughly $70$\,\% of $E_\mathrm{kin}$. We attribute this loss of heat to a combination of sources: uncertainty in the bolometer calibration, unaccounted heat exchange with the walls of the fused quartz cylinder, and low level of heat release at slow rate owing to a large relative share of the laminar response. Compared to the large and much shorter rectangular pulse, which an ideal turbulent front is expected to generate in steady motion according to Eq.~(\ref{dissi1}), the detected small and slow triangular signal is remarkably different.
\begin{figure}
\begin{center}
\includegraphics[%
  width=0.6\linewidth,
  keepaspectratio]{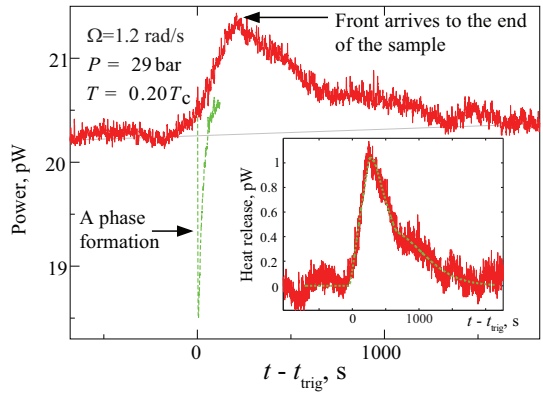}
\end{center}
\vspace{-5mm}
\caption{(color online) Thermal signal from the front propagation. The signal in the main panel illustrates the bolometer record where the front is triggered at $t=0$ by the creation of the A-phase layer (solid line). The cooling spike in the beginning caused by the absorbed latent heat of the A-B transition is subtracted using an identical cooling record measured in the equilibrium vortex state (dashed line). The inset shows the corresponding heat release from the front motion with the background heat leak of the order of 20~pW subtracted. The dashed line represents a fit to the thermal model discussed in the text.}
\label{signal}
\end{figure}

The substantial differences from the fully turbulent model of Eq.~(\ref{dissi1}) indicate that laminar flow figures importantly in the thermal response. Qualitatively, this is seen as the bump at $t~\sim1000$\,s and as the following slow decay of the thermal signal.
To quantify the significance of the laminar flow, let us use a simple model consisting of a turbulent front propagating
axially, followed by purely laminar motion, where the vortices extend in spiral configuration from the cylindrical wall inward towards the center. We assume solid-body rotation of the superfluid component everywhere and at all times, i.e., $\vs=\Oms(z-V_{\rm f} t) \hat{\bf z} \times \textbf{r}$. The steady-state profile of $\Oms(z)$ is organized as follows: In the front itself
$\Oms$ increases linearly from zero to $\epsilon \Omega$
over an axial length $w_\mathrm{f}$: $\Oms = \epsilon\Omega(1-z/w_{\rm f})$. The front is followed by a laminar
tail where the relaxation is given by
\begin{equation}
\label{up}
\Omega_{\mathrm{s}} (\tilde t)= \frac{\Omega}{1-(1-1/\epsilon)\exp (-\tilde
  t/\tau)}\,.
\end{equation}
Eq.~(\ref{up}) represents the spin-up of $\Omega_\mathrm{s}$ from the
initial velocity  $\epsilon \Omega$ to the drive velocity $\Omega$ by the
mutual-friction-assisted radial component of 2-dimensional spiraling vortex
motion in the azimuthal plane with the time constant $\tau (T) = [2 \,
\alpha (T) \, \Omega]^{-1}$.\cite{PRL10} This structure of vorticity is
pushed through the cylinder at a steady velocity $V_{\rm f}(\Omega, T)$ and thus $\tilde t = t - z/V_{\rm f}$. The total free energy of the superfluid is
calculated by substituting $\Oms(z,t)$ into Eq.~(\ref{freeen}) and
integrating over the length of the sample. Taking its negative derivative
over time gives the dissipation rate. The dashed line in the inset of
Fig. \ref{signal} is obtained by fitting the calculated dissipation
$\dot{Q} (t)$ to the measurement by means of the three parameters ($\epsilon$,
$w_\mathrm{f}$, and $V_{\rm f}$) using the measured bolometer rise time
$\tau_{\rm T}$ and the mutual-friction parameter $\alpha(T)$. \cite{vort_ann}

From the NMR measurements, we conclude that the peak of the thermal signal
corresponds to the arrival of the front to the end plate of the cylinder.
This fixes the front velocity $V_{\rm f}$. The narrow shoulder after the peak is a signature of the laminar relaxation, which limits $w_\mathrm{f}$. The division of the total energy dissipation in the turbulent front and in the laminar relaxation shows that the latter is 70\,\% in Fig.~\ref{signal} and the bundle behind the turbulent front contains only $\epsilon = 0.34$ of the equilibrium number of vortices. Thus, the superfluid behind the front is partially decoupled from the reference frame of the container.

The assumption that the superfluid is in solid-body-like rotation is a significant simplification. In reality, both the core of the front and the bundle behind it are twisted. Moreover, in the front vortices turn from the axial orientation to the plane perpendicular to the cylinder axis in order to end at the cylindrical boundary. In addition, comparison of the NMR signals measured when the front is within an NMR coil to the numerically-calculated spectra using model textures\cite{kopu} shows that the vortex configuration in the front is not perfectly uniform in the radial direction. Nonetheless, we believe that the obtained parameter values are reasonable estimates, which capture the essential physics.

\begin{figure}
\begin{center}
\includegraphics[%
  width=0.6\linewidth,
  keepaspectratio]{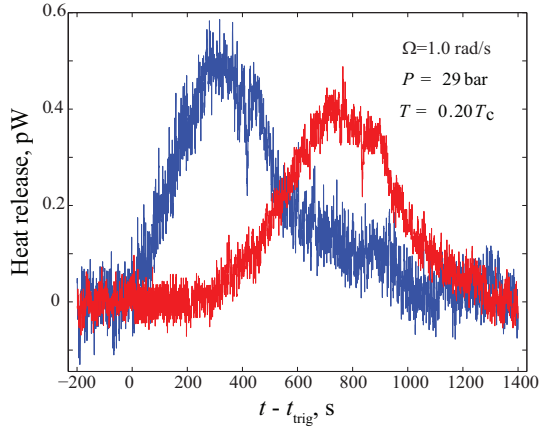}
\end{center}
\vspace{-5mm}
\caption{(color online) Effect of the rate at which the Kelvin-Helmholtz instability develops on the front propagation. In the case of the left curve, the front is triggered by decreasing the stabilization field much below the critical field $H\Sb{AB}$ at $t=0$, which leads to an instantaneous start of the heat release. For the right curve, the field is reduced barely below the critical field. In this case, the dynamics of the instability event is slower and the start of the front propagation and thus, the energy dissipation, is delayed by about 300~s. For the left curve $V_{\rm f}=0.0134$~mm/s and the fraction of the heat dissipated by the time the front reaches the end of the sample $Q_{\rm f}/Q_{\rm tot}=0.32$. For the right curve, the corresponding values are $V_{\rm f}=0.0124$~mm/s and $Q_{\rm f}/Q_{\rm tot}=0.28$. The initial heating spike from the fast change in the barrier-magnet field is subtracted from both signals.
}
\label{Q_comp}
\end{figure}
The characteristics of the thermal response depend on the initial vortex configuration and thus, on the triggering mechanism. Figure \ref{Q_comp} shows the responses from two experiments. For the left curve, the front is triggered in a rapid reduction of the stabilization field substantially below the critical field $H\Sb{AB}$, while for the right curve the field is decreased only slightly below $H\Sb{AB}$. In the former case, the instability develops instantaneously, while in the latter case the response is delayed by about 300~s and the travel time is slightly longer, similarly to the example of the NMR signal in the bottom part of Fig.~\ref{nmrsignal}. The A-phase-creation method always leads to an instantaneous start of the front propagation.

\section{Conclusions}
Both turbulent and laminar motion of quantized vortices in \he{3}-B lead to the creation of quasiparticle excitations. Our bolometric techniques enable the direct detection of these excitations with a resolution of 0.1~pW in a sample rotating at $\sim1$~rad/s. The thermal signal from the turbulent vortex-front motion reveals that at low temperatures, where the mutual-friction coupling is rapidly reduced with decreasing temperature, the superfluid in the front partially decouples from the reference frame of the container. As a result, a major fraction of the free-energy difference between the vortex-free state and the equilibrium array of vortices is dissipated in a laminar spin-up of the superfluid component.

The reduced superfluid angular velocity is supported by the vortex-line tension caused by the difference in the precession frequencies of the front and the twisted bundle behind it. The line tension counteracts the mutual-friction force, which drives the superfluid to corotate with the container. A quantitative analysis of the decoupling will be published elsewhere.

The thermal signal also allows us to study the dynamics of the Kelvin-Helmholtz instability. Depending on how rapidly the instability develops, the injection process of the vortices escaping from the A to the B phase varies. If the interface is strongly perturbed, a turbulent burst setting the front into motion is created instantaneously, while with a weaker perturbance, the response is delayed and the front propagates slower.


\begin{acknowledgements}
We thank Richard Haley, Risto H\"anninen, Victor L'vov, and Grigori Volovik for collaboration and discussions. The work is supported by the Academy of Finland and the EU 7th Framework Programme (FP7/2007-2013, grant 228464 Microkelvin). J.H. acknowledges financial support from the V\"{a}is\"{a}l\"{a} Foundation of the Finnish Academy of Science and Letters.
\end{acknowledgements}

\bibliographystyle{spphys}
\bibliography{references}   

\begin{thebibliography}{10}
\providecommand{\url}[1]{{#1}}
\providecommand{\urlprefix}{URL }
\expandafter\ifx\csname urlstyle\endcsname\relax
  \providecommand{\doi}[1]{DOI \discretionary{}{}{}#1}\else
  \providecommand{\doi}{DOI \discretionary{}{}{}\begingroup
  \urlstyle{rm}\Url}\fi

\bibitem{front1}
V.B. Eltsov, A.I. Golov, R.~de~Graaf, R.~H\"anninen, M.~Krusius, V.S. L'vov,
  R.E. Solntsev, Phys. Rev. Lett. \textbf{99}, 265301 (2007)

\bibitem{donnelly}
R.J. Donnelly, \emph{Quantized Vortices in Helium II} (Cambridge University
  Press, Cambridge, 1991)

\bibitem{kopnin}
N.B. Kopnin, Rep. Prog. Phys. \textbf{65}, 1633 (2002)

\bibitem{bevan}
T.D.C. Bevan, A.J. Manninen, J.B. Cook, A.J. Armstrong, J.R. Hook, H.E. Hall,
  Phys. Rev. Lett. \textbf{74}, 750 (1995)

\bibitem{volovik04}
G.E. Volovik, J. Low Temp. Phys. \textbf{136}, 309 (2004)

\bibitem{PRL10}
V.B. Eltsov, R.~de~Graaf, P.J. Heikkinen, J.J. Hosio, R.~H\"anninen,
  M.~Krusius, V.S. L'vov, Phys. Rev. Lett. \textbf{105}, 125301 (2010)

\bibitem{Manchester}
P.M. Walmsley, A.I. Golov, H.E. Hall, A.A. Levchenko, W.F. Vinen, Phys. Rev.
  Lett. \textbf{99}, 265302 (2007)

\bibitem{AB_turb}
P.M. Walmsley, V.B. Eltsov, P.J. Heikkinen, J.J. Hosio, R.~H\"anninen,
  M.~Krusius, Phys. Rev. B \textbf{84}, 184532 (2011)

\bibitem{prec}
J.J. Hosio, V.B. Eltsov, M.~Krusius, J.T. M\"akinen, Phys. Rev. B \textbf{85},
  224526 (2012)

\bibitem{vinen}
W.F. Vinen, J. Low Temp. Phys. \textbf{161}, 419 (2010)

\bibitem{vinenprb}
W.F. Vinen, Phys. Rev. B \textbf{64}, 134520 (2001)

\bibitem{silaev}
M.A. Silaev, Phys. Rev. Lett. \textbf{108}, 045303 (2012)

\bibitem{front2}
J.J. Hosio, V.B. Eltsov, R.~de~Graaf, P.J. Heikkinen, R.~H\"anninen,
  M.~Krusius, V.S. L'vov, G.E. Volovik, Phys. Rev. Lett. \textbf{107}, 135302
  (2011)

\bibitem{lanc_nat}
D.I. Bradley, S.N. Fisher, A.M. Gu\'enault, R.P. Haley, G.R. Pickett, D.~Potts,
  V.~Tsepelin, Nature Phys. \textbf{7}, 473 (2011)

\bibitem{PLTP_rota}
V.B. Eltsov, R.~de~Graaf, R.~H\"anninen, M.~Krusius, R.E. Solntsev, V.S. L'vov,
  A.I. Golov, P.~Walmsley, Prog. Low Temp. Phys. \textbf{XVI}, 45 (2009)

\bibitem{finne}
A.P. Finne, V.B. Eltsov, R.~Blaauwgeers, Z.~Janu, M.~Krusius, L.~Skrbek, J. Low
  Temp. Phys. \textbf{134}, 375 (2004)

\bibitem{Nature03}
A.P. Finne, T.~Araki, R.~Blaauwgeers, V.B. Eltsov, N.B. Kopnin, M.~Krusius,
  L.~Skrbek, M.~Tsubota, G.E. Volovik, Nature \textbf{424}, 1022 (2003)

\bibitem{precursor}
A.P. Finne, V.B. Eltsov, G.~Eska, R.~H\"anninen, J.~Kopu, M.~Krusius, E.V.
  Thuneberg, M.~Tsubota, Phys. Rev. Lett. \textbf{96}, 085301 (2006)

\bibitem{twist_prl}
V.B. Eltsov, A.P. Finne, R.~H\"anninen, J.~Kopu, M.~Krusius, M.~Tsubota, E.V.
  Thuneberg, Phys. Rev. Lett. \textbf{96}, 215302 (2006)

\bibitem{twist_jltp}
V.B. Eltsov, R.~de~Graaf, R.~H\"anninen, M.~Krusius, R.E. Solntsev, J. Low
  Temp. Phys. \textbf{150}, 373 (2008)

\bibitem{sonin_front}
E.B. Sonin, Phys. Rev. B \textbf{85}, 024515 (2012)

\bibitem{karimaki}
J.M. Karim\"aki, R.~H\"anninen, E.V. Thuneberg, Phys. Rev. B \textbf{85},
  224519 (2012)

\bibitem{VorFormAnnih}
V.B. Eltsov, R.~de~Graaf, P.J. Heikkinen, J.J. Hosio, R.~H\"anninen,
  M.~Krusius, J. Low Temp. Phys. \textbf{161}, 474 (2010)

\bibitem{KH_instab}
R.~Blaauwgeers, V.B. Eltsov, G.~Eska, A.P. Finne, R.P. Haley, M.~Krusius, J.J.
  Ruohio, L.~Skrbek, G.E. Volovik, Phys. Rev. Lett. \textbf{89}, 155301 (2002)

\bibitem{deGraaf}
R.~de~Graaf, V.B. Eltsov, P.J. Heikkinen, J.J. Hosio, M.~Krusius, J. Low Temp.
  Phys. \textbf{163}, 238 (2011)

\bibitem{magnon}
S.~Autti, Y.M. Bunkov, V.B. Eltsov, P.J. Heikkinen, J.J. Hosio, P.~Hunger,
  M.~Krusius, G.E. Volovik, Phys. Rev. Lett. \textbf{108}, 145303 (2012)

\bibitem{andreev_hosio}
J.J. Hosio, V.B. Eltsov, R.~de~Graaf, M.~Krusius, J.~M\"akinen, D.~Schmoranzer,
  Phys. Rev. B \textbf{84}, 224501 (2011)

\bibitem{todo}
I.A. Todoschenko, A.~Alles, H.~Babkin, A.Y. Parshin, V.~Tsepelin, J. Low Temp.
  Phys. \textbf{126}, 1446 (2002)

\bibitem{PhysRevLett.83.3462}
M.~Bartkowiak, S.W.J. Daley, S.N. Fisher, A.M. Gu\'enault, G.N. Plenderleith,
  R.P. Haley, G.R. Pickett, P.~Skyba, Phys. Rev. Lett. \textbf{83}, 3462 (1999)

\bibitem{Abar_shape}
V.B. Eltsov, R.~Blaauwgeers, A.P. Finne, M.~Krusius, J.J. Ruohio, G.E. Volovik,
  Physica B \textbf{329-333}, 96 (2003)

\bibitem{PhysRevLett.85.4321}
M.~Bartkowiak, S.N. Fisher, A.M. Gu\'enault, R.P. Haley, G.R. Pickett, G.N.
  Plenderleith, P.~Skyba, Phys. Rev. Lett. \textbf{85}, 4321 (2000)

\bibitem{vort_ann}
V.B. Eltsov, R.~de~Graaf, P.J. Heikkinen, J.J. Hosio, R.~H\"anninen,
  M.~Krusius, J. Low Temp. Phys. \textbf{161}, 474 (2010)

\bibitem{kopu}
J.~Kopu, J. Low Temp. Phys. \textbf{146}, 47 (2007)

\end{thebibliography}
\end{document}